\documentclass[12pt, prd, showpacs]{revtex4}
\usepackage{amssymb}
\usepackage{amsmath}

\input{tcilatex}

\begin{document}

\title{Energetics of particle collisions near dirty rotating extremal black
holes:\\
Banados-Silk-West effect versus Penrose process}
\author{O. B. Zaslavskii}
\affiliation{Department of Physics and Technology, Kharkov V.N. Karazin National
University, 4 Svoboda Square, Kharkov, 61077, Ukraine}
\email{zaslav@ukr.net }

\begin{abstract}
If two particles collide near the horizon of a rotating extremal black hole,
under certain conditions the energy $E_{c.m.}$ in the center-of-mass frame
can grow without limit (the so-called Banados-Silk-West effect). We consider
collisions that produce two other particles. We show that for a generic
dirty (surrounded by matter) black hole, there exist upper bounds on the
energy and mass of product particles which can be detected at infinity. As a
result, the positive energy gain is possible but is quite modest. It mainly
depends on two numbers in which near-horizon behavior of the metric is
encoded. The obtained results suggest astrophysical limits on the
possibility of observation of the products of the collisional
Banados-Silk-West effect. These results are consistent with recent
calculations for the Kerr metric, extending them to generic dirty black
holes. It is shown that for dirty black holes there are types of scenarios
of energy extraction impossible in the Kerr case..
\end{abstract}

\keywords{black hole horizon, centre of mass, acceleration of particles}
\pacs{04.70.Bw, 97.60.Lf }
\maketitle

\section{Introduction}

Recently, the effect of unbound energy in the centre of mass frame of
particles colliding near black holes was discovered \cite{ban} ((called the
Banados-Silk-West (BSW) effect after the names of its authors).). This
effect is interesting from the theoretical point of view since new physics
can come into play in the vicinity of black holes at the Planck scale and
beyond it, with new channels of reactions between particles. From the other
hand, there are potential astrophysical manifestations of the BSW effect.
Some of them can take place in the vicinity of the horizon, i.e. just the
region where this effect occurs. This includes, for example, physics of
accretion disks and behavior of extreme mass-ratio inspirals \cite{ac} - 
\cite{spiral}.

Meanwhile, there are also observations on Earth which could be supposedly
interpreted on the basis of the BSW effect. In particular, it was
conjectured that ultra-high energy cosmic rays detected by the AUGER group 
\cite{aug} might be created in the active galactic nuclei due to the BSW
effect in the vicinity of the central supermassive black hole \cite{gp} - 
\cite{gpp}. Quite recently, another possible astrophysical manifestation of
the BSW effect connected with the processes with neutralino was discussed in 
\cite{n}. For such kind of observations, it is important to know what masses
and energies of particles can be detected at infinity. To this end, in \cite%
{flux1}, \cite{flux2} the emergent flux emitted by dark matter spikes around
intermediate-mass black hole was evaluated from the region close to the
horizon. We choose a different approach and consider elementary acts of
collisions when a pair of particles is converted into a pair of two other
ones. We consider the process in the immediate vicinity of the horizon only
that enables us to analyze the problem from the first principles.

The first predictions of such a kind were made in \cite{ted} where the
authors claimed the existence of the bound on the ratio $E/m,$ where $E$ is
the energy of the particle detected at infinity and $m$ is the mass of
infalling particles. The result was criticized in \cite{gp} where it was
noticed that the deviation of \cite{ted} was made neglecting the difference
between the time-like four-velocity vector and the light-like horizon
generators. However, the numeric example suggested in \cite{gp} for the case
of the Kerr metric as a counter-example to \cite{ted}, applies to the
collision not on the horizon but at some point outside its immediate
vicinity, so it cannot be related to the BSW effect directly. Therefore, the
issue discussed in \cite{gp} remained incomplete and more general treatment
is needed to conclude whether or not the bounds on the energy of products of
the BSW effect exist.

Quite recently, the problem was considered anew and it was pointed in works 
\cite{p} - \cite{j} that upper limits on the energy of particles detected at
infinity do exist. This was done in \cite{p}, \cite{j} for the Kerr metric.
In \cite{z1}, the problem was studied for generic "dirty" (surrounded by
matter) black holes but, in particular, one important scenario was
overlooked. The aim of the present paper is to suggest a general complete
analysis of the reaction between two particles in the context of the BSW\
process and derive bounds on the energy and mass of particles escaping to
infinity applicable to generic dirty black holes. The last point is
especially important since in real astrophysical conditions matter is always
present near the horizon of a black hole.

The BSW effect was originally discovered for extremal black holes \cite{ban}%
. It exists also for nonextremal black holes \cite{gp}, but requires special
conditions like multiple scattering (see also \cite{prd} for
generalization). To avoid these subtleties not connected with the issue
under discussion directly, in the present paper we restrict ourselves to
extremal black holes. While analyzing the products of the BSW effect
detected at infinity, we make focus on the question whether or not one can
gain more energy than it was injected. In other words, we discuss the
possibility and the limitations of the Penrose process \cite{pen} in such
situation.

The BSW effect implies that both colliding particles move towards a black
hole and, additionally, parameters of one particle are fine-tuned. It should
not be confused with a more simple effect which arises when one of the
colliding particles moves towards the horizon and the other one moves away
from it \cite{ps}. We call the latter the Piran and Shanam (PS) effect, this
word is used here in the sense of physical effect irrespective of its
relevance or irrelevance in practical astrophysics \cite{ps2}. In the PS
effect, the energy in the centre of mass frame grows unbounded just due to
the blue-shift of energy, without any fine-tuning. Kinematically, in the
case of the PS process, particles experience head-on collision with at least
one of them having the speed almost equal to that of light in the frame of a
stationary observer. As a result, the relative velocity also tends to the
speed of light and the corresponding energy grows indefinitely. Near
rotating black holes, the PS effect can lead to the Penrose process (details
of the Penrose process in the background of charged and rotating black holes
can be found in \cite{nar}). From the other hand, for the BSW effect to
occur, both particles have to approach the horizon and special conditions
are required to achieve the relative velocity which would approach the speed
of light \cite{k}. There is also difference between the BSW and PS effects
in geometric terms \cite{cqg}.

In the present paper, we combine the approaches of \cite{j} and \cite{z1},
thus generalizing the results of \cite{p} and \cite{j}, derived for the Kerr
metric, to a generic dirty rotating extremal black hole. The details of this
presentation run along the lines of \cite{j} and at each step the emphasis
is made on features that are absent in the Kerr case and arise due to
"dirtiness" of black holes.

The general structure of the paper is the following. In Sec. II, we write
down the conservation laws of the radial and angular momenta and energy in
the act of collision. We also trace how this is related to the expansion of
the four-velocity in terms of the local null tetrad. Division of all
particles to two classes (critical and usual) is introduced which is crucial
for what follows. In Sec. III, the general conditions are formulated under
which a particle can escape to infinity. In Sec. IV, we write down the
expansions of the metric coefficient and radial momenta near the horizon in
terms of the lapse function $N,$ which is a small quantity near the horizon.
This is done separately for usual and (near)critical particles. In Sec. V,
general classification of scenarios of collision is suggested according to
the kind of particles (usual or critical) and the direction of their motion
immediately after collision. In Sec. VI, the conservation of radial momentum
is analyzed in the first order in $N$ using the results of Sec. IV. In Sec.
VII the same procedure is carried out with terms of the order $N^{2}$ taken
into account. In Sec. VIII, all allowed scenarios of collision are analyzed
and the upper bounds on the mass and energy of particles escaping to
infinity are obtained for each scenario separately. This is obtained from
the conditions of escaping derived in Sec. III and the conservation law for
the radial momentum analyzed in Sec. VI\ and VII. In Sec. IX, these results
are used to elucidate whether energy extraction is possible. Again, all the
scenarios are analyzed. In Sec. X, some concrete types of reactions are
considered as illustration. The main results are summarized in Sec. XI. In
two Appendices we list some technical details used in the main text.

We use units in which fundamental constants $G=c=$%
h{\hskip-.2em}\llap{\protect\rule[1.1ex]{.325em}{.1ex}}{\hskip.2em}%
$=1.$

\section{Equations of motion and conservation laws}

Let us consider the axially symmetric black hole metric 
\begin{equation}
ds^{2}=-N^{2}dt^{2}+g_{\phi \phi }(d\phi -\omega dt)^{2}+dn^{2}+g_{zz}dz^{2}%
\text{,}
\end{equation}%
where the metric coefficients do not depend on $t$ and $\phi $. We want to
write down the conservation laws in the act of collision for particles
moving in this background. First of all, we must relate the characteristics
of each particle to the properties of the metric. Let a particle having the
four-velocity $u^{\mu }$ move in this background. It is convenient to expand 
$u^{\mu }$ with respect to the null tetrad basis. We can write 
\begin{equation}
u^{\mu }=\frac{l^{\mu }}{2\alpha }+\beta N^{\mu }+s^{\mu }  \label{ur}
\end{equation}%
where $l^{\mu }$ and $N^{\mu }$ are null vectors normalized according to $%
l_{\mu }N^{\mu }=-1$, $s^{\mu }$ is a space-like vector orthogonal to $%
l^{\mu }$ and $N^{\mu }$, $s^{\mu }=s_{a}a^{\mu }+s_{b}b^{\mu }$ where $%
a^{_{\mu }}$ and $b^{\mu }$ are unit vectors orthogonal to each other and to 
$l^{\mu }$ and $N^{\mu }$. Then, we can choose%
\begin{equation}
l^{\mu }=(1,N,\omega ,0)\text{,}  \label{lw}
\end{equation}%
\begin{equation}
N^{\mu }=\frac{1}{2}(\frac{1}{N^{2}},-\frac{1}{N},\frac{\omega }{N^{2}},0)
\label{nw}
\end{equation}%
where $x^{\mu }=(t,n,\phi ,z)$.

One can check that the decomposition of the metric%
\begin{equation}
g_{\alpha \beta }=-l_{\alpha }N_{\beta }-l_{\beta }N_{\alpha }+a_{\alpha
}a_{\beta }+b_{\alpha }b_{\beta }
\end{equation}%
is satisfied with%
\begin{equation}
a^{\mu }=(0,0,\frac{1}{\sqrt{g_{\phi \phi }}},0)\text{, }
\end{equation}%
\begin{equation}
b^{\mu }=\frac{1}{\sqrt{g_{zz}}}(0,0,0,1)\text{, }b^{\mu }b_{\mu }=a^{\mu
}a_{\mu }=1\text{,}
\end{equation}%
\begin{equation}
\alpha =\frac{\beta }{\delta }\text{, }\delta =1+s^{2}\text{, }s^{2}=s^{\mu
}s_{\mu }\text{.}  \label{abd}
\end{equation}

We assume that the metric is symmetric with respect to $z$ and restrict
ourselves by the motion within the equatorial plane. Then, 
\begin{equation}
s_{a}\equiv u_{\mu }a^{\mu }=\frac{L}{\sqrt{g}}\text{, }s_{b}\equiv u^{\mu
}b_{\mu }=0\text{, }s^{2}=\frac{L^{2}}{m^{2}g}\text{,}
\end{equation}%
where $L=mu_{\phi }$ is angular momentum per unit mass that conserves due to
the independence of the metric on $\phi $. For shortness, we use the
notation $g\equiv g_{\phi \phi }$. In a \ similar way, $u_{0}$ is conserved
because of the independence of the metric on $t$, $E=-mu_{0}$ being the
energy. Expressing $u^{\mu }$ in terms of $u_{\mu }$ and using (\ref{ur})
and the normalization conditions $u^{\mu }u_{\mu }=-1$, one obtains that 
\begin{equation}
mu^{0}=\frac{X}{N^{2}}\text{, }
\end{equation}%
\begin{equation}
mu^{3}=\frac{L}{g}+\frac{\omega X}{N^{2}}
\end{equation}%
and%
\begin{equation}
mu^{1}=\varepsilon \frac{Z}{N}\text{,}  \label{u1}
\end{equation}%
\begin{equation}
X=E-\omega L\text{,, }Z=\sqrt{X^{2}-N^{2}(m^{2}+\frac{L^{2}}{g})},
\label{zu}
\end{equation}%
where $\varepsilon =+1$ for an outgoing particle and $\varepsilon =-1$ for
an ingoing one. 

Then, it is easy to calculate%
\begin{equation}
\frac{1}{2\alpha }=-u_{\mu }N^{\mu }\text{, }\alpha =\frac{mN^{2}}{%
X+\varepsilon Z}\text{, }\beta =-u_{\mu }l^{\mu }=\frac{X-\varepsilon Z\text{%
.}}{m}\text{. }  \label{ab}
\end{equation}%
One can check that the coefficients $\alpha $ and $\beta $ obey Eq.(\ref{abd}%
). We also assume the forward in time condition $u^{0}>0$ as usual, so $X>0$
everywhere except, possibly, on the horizon where $X=0$ is also allowed. It
is seen from (\ref{zu}) that $Z\leq X$. Therefore, $\alpha $, $\beta \geq 0$%
. If the horizon value $X_{H}$ is positive we call a particle usual. If $%
\left( X\right) _{H}=0$ it is called critical (near-critical if $\left(
X\right) _{H}$ is small).

Equating the coefficients at $l_{\mu }$, $N_{\mu }$ and $a^{\mu }$ we obtain
for the reaction when two initial particles turn into particles with masses $%
m_{3}$ and $m_{4}$:

\begin{equation}
m_{1}\beta _{1}+m_{2}\beta _{2}=m_{3}\beta _{3}+m_{4}\beta _{4}\text{,}
\label{ba}
\end{equation}%
\begin{equation}
\frac{m_{1}}{\alpha _{1}}+\frac{m_{2}}{\alpha _{2}}=\frac{m_{3}}{\alpha _{3}}%
+\frac{m_{4}}{\alpha _{4}}\text{,}  \label{1ar}
\end{equation}%
\begin{equation}
L_{1}+L_{2}=L_{3}+L_{4}.  \label{l12}
\end{equation}

Equivalently, one can find from (\ref{ba}) - (\ref{l12}) that%
\begin{equation}
E_{1}+E_{2}=E_{3}+E_{4}\text{,}  \label{e}
\end{equation}%
\begin{equation}
X_{1}+X_{2}=X_{3}+X_{4}\text{,}  \label{x}
\end{equation}%
\begin{equation}
\varepsilon _{1}Z_{1}+\varepsilon _{2}Z_{2}=\varepsilon
_{3}Z_{3}+\varepsilon _{4}Z_{4}  \label{ez}
\end{equation}%
where Eq. (\ref{ez}) has the meaning of the conservation of the radial
momentum.

It is worth stressing that individual energies $E_{i}$ are finite. It is the
energy in the centre of mass frame $E_{c.m.}$ which is divergent if the BSW
effect takes place (see, e.g. \cite{prd} for details).

\section{Conditions of escaping to infinity}

As $u^{1}\sim Z$ according to (\ref{u1}), the zeros of $Z$ give us the
turning points. The condition $Z=0$ can be rewritten as%
\begin{equation}
l^{2}-2\frac{\omega le}{(\omega ^{2}-\frac{N^{2}}{g})}+\frac{(e^{2}-N^{2})}{%
(\omega ^{2}-\frac{N^{2}}{g})}=0\text{, }e=\frac{E}{m}\text{, }l=\frac{L}{m}%
\text{,}
\end{equation}%
whence its roots equal%
\begin{equation}
l_{\pm }=\frac{e\pm N\sqrt{Y}}{(\omega -\frac{N^{2}}{g\omega })}\text{, }Y=%
\frac{e^{2}+g_{00}}{g\omega ^{2}}\text{, }g_{00}=-N^{2}+g\omega ^{2}\text{.}
\label{lt}
\end{equation}%
On the horizon, 
\begin{equation}
L_{\pm }\equiv ml_{\pm }=\frac{E}{\omega _{H}}=L_{H}.
\end{equation}

The allowed region of motion corresponds to $Z^{2}\geq 0$ where $Z$ is given
by Eq. (\ref{zu}). The region between turning points is forbidden. We will
be interested in the situation when a particle (denoted as particle 3)
escapes to infinity from the immediate vicinity of the horizon. This is
possible in 2 cases that generalizes the corresponding situation in the Kerr
metric \cite{j}.

a) $E_{3}>m_{3}$, $L_{3}<L_{H}$, $\varepsilon _{3}=+1$. b) $E_{3}\geq m_{3}$%
, $L_{H}<L_{3}<L_{-}(E_{3})$, $\varepsilon _{3}=+1$ or $\varepsilon _{3}=-1$%
. The condition $\varepsilon _{3}=-1$ means in this context that particle 3
is moving inward, approaches the outer turning point and bounces back in the
outward direction. We consider all these types of scenario in the vicinity
of the horizon where $N\ll 1$.

It is convenient to write%
\begin{equation}
L=\frac{E}{\omega _{H}}(1+\delta ).  \label{ld}
\end{equation}

Then, in case (a) 
\begin{equation}
\delta <0.  \label{dneg}
\end{equation}%
In case (b) 
\begin{equation}
\delta \geq 0  \label{dpos}
\end{equation}%
but it is bounded from above. Indeed, forward in time condition $X=E-\omega
L>0$ gives us%
\begin{equation}
\delta <\frac{\omega _{H}-\omega }{\omega }\text{.}  \label{d}
\end{equation}

\section{Near-horizon expansions}

In this region, the lapse function $N$ \ is a small quantity and the
expansion of the coefficient $\omega $ near the extremal horizon takes the
general form 
\begin{equation}
\omega =\omega _{H}-B_{1}N+B_{2}N^{2}+O(N^{3})  \label{om}
\end{equation}%
where $\omega _{H}$ is the horizon value of $\omega $ and $B_{i}$ is some
model-dependent coefficient \cite{t}. Also, we will use the expansion for
the metric coefficient $g$%
\begin{equation}
g=g_{H}+g_{1}N+g_{2}N^{2}+O(N^{3})\text{.}  \label{g}
\end{equation}

For what follows, we need also the expansions for the quantity $Z$. This can
be found separately for different kinds of particles.

\subsection{Usual particle}

For such \ a particle, $X_{H}\neq 0$, so we obtain

\begin{equation}
Z=X-\frac{1}{2X}(m^{2}+\frac{L^{2}}{g_{H}})N^{2}+O(N^{3})  \label{uz}
\end{equation}%
where%
\begin{equation}
X=X_{H}+B_{1}LN-B_{2}LN^{2}+...
\end{equation}

\subsection{Critical particle}

Now, $X_{H}=0$, $L=\frac{E}{\omega _{H}}$, so%
\begin{equation}
X=\frac{E}{\omega _{H}}(\omega _{H}-\omega )=\frac{EN}{h}(b-b_{2}N)+O(N^{3})
\label{cx}
\end{equation}%
\begin{equation}
Z=\sqrt{E^{2}\frac{(b^{2}-1)}{h^{2}}-m^{2}}N+\frac{E^{2}}{h^{2}}\frac{(\frac{%
g_{1}}{2g_{H}}-bb_{2})}{\sqrt{E^{2}\frac{(b^{2}-1)}{h^{2}}-m^{2}}}%
N^{2}+O(N^{3}),  \label{cz}
\end{equation}%
where we introduced useful notations%
\begin{equation}
b=B_{1}\sqrt{g_{H}}\text{, }h=\omega _{H}\sqrt{g_{H}}\text{, }b_{2}=B_{2}%
\sqrt{g_{H}}\text{.}  \label{bh}
\end{equation}

Hereafter, we assume that $B_{1}>0$ to satisfy the forward in time condition 
$X>0$.

\subsection{Near-critical particle}

Let us consider a particle which is not exactly critical but, rather,
near-critical. It has the angular momentum (\ref{ld}) with $\delta \ll 1$.
Then, it follows from (\ref{d}) and (\ref{om}) that%
\begin{equation}
\delta <bN+N^{2}(b^{2}-b_{2})+O(N^{3}).  \label{dn}
\end{equation}%
On the horizon, $\left( X_{3}\right) _{H}=-\frac{E_{3}}{\omega _{H}}\delta $.

Near the horizon, we can take $\delta $ that would adjust to small value of $%
N$ and write the expansion%
\begin{equation}
\delta =C_{1}N+C_{2}N^{2}+...  \label{de}
\end{equation}

Then, 
\begin{equation}
X=NE(\frac{b}{h}-C_{1})+QEN^{2}+O(N^{3}),  \label{nx}
\end{equation}%
\begin{equation}
Q=C_{1}\frac{b}{h}-\frac{b_{2}}{h}-C_{2},
\end{equation}%
\begin{equation}
Z=N\sqrt{E^{2}[(\frac{b}{h}-C_{1})^{2}-\frac{1}{h^{2}}]-m^{2}}+\tau
N^{2}+O(N^{3}),  \label{nz}
\end{equation}%
\begin{equation}
\tau =\frac{E^{2}(\rho +\frac{1}{2h^{2}}\frac{g_{1}}{g_{H}})}{\sqrt{E^{2}[(%
\frac{b}{h}-C_{1})^{2}-\frac{1}{h^{2}}]-m^{2})}},  \label{ga}
\end{equation}%
\begin{equation}
\rho =-C_{1}^{2}\frac{b}{h}+C_{1}C_{2}+C_{1}\left( \frac{b^{2}-1}{h^{2}}+%
\frac{b_{2}}{h}\right) -C_{2}\frac{b}{h}-\frac{bb_{2}}{h^{2}}.  \label{ro}
\end{equation}

\subsection{Comparison with the case of the extremal Kerr black hole}

For what follows we need the corresponding quantities for the simplest case
of the Kerr extremal metric. This will enable us to compare the general
formulas with the results of \cite{p}, \cite{j}. We are considering the
plane $\theta =\frac{\pi }{2}$ where for the Kerr metric

\begin{equation}
N^{2}=\frac{(r-M)^{2}}{r^{2}+M^{2}+\frac{2M^{3}}{r}},
\end{equation}

\begin{equation}
\omega =\frac{2M^{2}}{r^{3}+M^{2}r+2M^{3}}\text{,}
\end{equation}%
\begin{equation}
g=r^{2}+M^{2}+\frac{2M^{3}}{r}\text{.}
\end{equation}%
Then, in the expansions (\ref{om}), (\ref{g})%
\begin{equation}
\omega _{H}=\frac{1}{2M}\text{, }B_{1}=\frac{1}{M}\text{, }B_{2}=\frac{1}{2M}%
\text{, }g_{H}=4M^{2}\text{, }g_{1}=0\text{.}  \label{kerr}
\end{equation}%
\begin{equation}
b=2\text{, }h=1\text{, }b_{2}=1.  \label{bbh}
\end{equation}

In \cite{j} the authors used the expansion in the form $\delta =\delta
_{1}\varepsilon +\delta _{2}\varepsilon ^{2}+...$ where near the horizon $r=%
\frac{M}{1-\varepsilon }$ with $\varepsilon \ll 1$.

One should bear in mind that the quantity $N\approx \frac{\varepsilon }{2}+%
\frac{\varepsilon ^{2}}{2}+O(\varepsilon ^{3})$. Correspondingly,%
\begin{equation}
C_{1}=2\delta _{1}\text{, }C_{2}=4\delta _{2}-4\delta _{1}\text{.}
\label{cd}
\end{equation}

Then,%
\begin{equation}
\rho =-2+8(2\delta _{1}-\delta _{2}-2\delta _{1}^{2}+\delta _{1}\delta _{2})%
\text{.}  \label{rok}
\end{equation}%
\newline
Eq. (\ref{dn}) turns into $\delta <\varepsilon +\frac{7}{4}\varepsilon
^{2}+O(\varepsilon ^{3})$ in agreement with Eq. (3.11) of \cite{j}.

\section{Allowed scenarios of collisions}

We assume that particles 1 and 2 move towards the horizon, so $\varepsilon
_{1}=\varepsilon _{2}=-1$. The BSW effect is possible only if one of them is
critical whereas the other one is usual. Let particle 1 be critical. Then,
we must consider different situations with $\varepsilon _{3}$ and $%
\varepsilon _{4}$ using the momentum conservation (\ref{ez}). Some
information can be extracted from the calculation of the left and right hand
sides of that equation on the horizon. The left hand side of (\ref{ez}) is
negative there.

1) $\varepsilon _{3}=\varepsilon _{4}$ \ It follows from (\ref{x}) that the
right hand side of (\ref{ez}) is equal to $\varepsilon _{3}\left(
X_{2}\right) _{H}$. As $\left( X_{2}\right) _{H}>0$, we must have $%
\varepsilon _{3}=\varepsilon _{4}=-1$.

If both particles 3 and 4 are critical, Eq. (\ref{x}) is not satisfied on
the horizon since the right hand side is equal to zero whereas the left hand
side is not. One can also see that they cannot be both usual. This follows
from the asymptotic behavior (\ref{uz}) and (\ref{cz}) for usual and
critical particles. Namely, there are terms of the order $N$ in the left
hand side which do not have counterparts in the right hand side. The
conclusion is that either particle 3 is critical, particle 4 is usual or
vice versa. Below, neglecting in the main approximation terms containing $%
\delta $ and $N$ we obtain the following possibilities.

2) $\varepsilon _{3}=-\varepsilon _{4}$, Then, it follows from (\ref{x})
that the right hand side of Eq. (\ref{ez}) is equal to%
\begin{equation}
\varepsilon _{3}[2\left( X_{3}\right) _{H}-\left( X_{2}\right) _{H}].
\end{equation}

a) $\varepsilon _{3}=1$, $\varepsilon _{4}=-1$. Comparing with the left hand
side of (\ref{x}) \ we obtain that%
\begin{equation}
\left( X_{3}\right) _{H}=0\text{, }
\end{equation}

so particle 3 is critical$.$ Then, (\ref{x}) also tells us that

\begin{equation}
\left( X_{4}\right) _{H}=\left( X_{2}\right) _{H}>0\text{, }  \label{x24}
\end{equation}%
so particle 4 is usual.

b) $\varepsilon _{3}=-1$, $\varepsilon _{4}=+1$. In the same manner, we
obtain that%
\begin{equation}
\left( X_{3}\right) _{H}=\left( X_{2}\right) _{H}>0\text{, }
\end{equation}%
\begin{equation}
\left( X_{4}\right) _{H}=0\text{,}
\end{equation}%
so particle 3 is usual, particle 4 is critical.

Thus in the pair of particles 3 and 4 it is just the particle escaping to
infinity which is critical. For definiteness, we assume that it is particle
3. Then, it follows from the previous analysis that 
\begin{equation}
\varepsilon _{1}=\varepsilon _{2}=\varepsilon _{4}=-1  \label{eps}
\end{equation}%
in all cases. If collision occurs not exactly on the horizon but in its
vicinity (as it happens for relevant scenarios - see below), particle 3 is
not precisely critical but near-critical, so Eq. (\ref{ld}) holds for it.
Then, it follows from (\ref{x}) that in the main approximation (\ref{x24})
is satisfied.

Now, we can apply the near-horizon expansion to different scenarios of
escaping. In case (a), Eqs. (\ref{dneg}) and (\ref{de}) give us%
\begin{equation}
C_{1}<0\text{.}  \label{cneg}
\end{equation}

In case (b), we must take into account the presence of the turning point
outside the horizon. Then, expanding (\ref{lt}) and neglecting the terms of
the second order and higher, we obtain that 
\begin{equation}
0\leq C_{1}\leq \left( C_{1}\right) _{m}=\frac{b}{h}-\sqrt{\frac{m_{3}^{2}}{%
E_{3}^{2}}+\frac{1}{h^{2}}}.  \label{cc}
\end{equation}

The scenarios in which a near-critical particle has $\varepsilon _{3}=-1$
immediately after collision and thus moves inward will be called IN
scenarios for shortness. If after collision $\varepsilon _{3}=+1$ we will
call it "OUT" scenario. In turn, we will add "$-$" if $\delta <0$ and "+" if 
$\delta \geq 0$. In other words, we enumerate possible types of scenarios
characterizing them by signs of two quantities - $\varepsilon $ and $\delta $%
. In general, there are 4 combinations: OUT$-$, OUT$+$, IN$+$ and IN$^{\_}$.
However, it follows from (\ref{dpos}) that the scenario IN$-$ should be
rejected. For the Kerr metric, the types OUT$-$ and OUT$+$ are uninteresting
since they do not allow energy extraction \cite{p}, \cite{j}. However, this
is not necessarily so for dirty black holes, so we must discuss all the
three remaining types of scenarios.

\section{Conservation of momentum in first order}

Now, we have the situation in which particle 1 is critical, particles 2 and
4 are usual, particle 3 is near-critical. Equating terms of the first order
in $N$ in (\ref{ez}), and using (\ref{uz}), (\ref{cx}), (\ref{cz}), (\ref{nx}%
), (\ref{nz}), (\ref{ga}),(\ref{eps}) we obtain%
\begin{equation}
F\equiv A_{1}+E_{3}(C_{1}-\frac{b}{h})=\varepsilon _{3}\sqrt{E_{3}^{2}[(%
\frac{b}{h}-C_{1})^{2}-\frac{1}{h^{2}}]-m_{3}^{2}}  \label{1}
\end{equation}%
where%
\begin{equation}
A_{1}=\frac{E_{1}b-\sqrt{E_{1}^{2}(b^{2}-1)-m_{1}^{2}h^{2}}}{h}.  \label{a1b}
\end{equation}

Taking here the square of (\ref{1}) we get%
\begin{equation}
C_{1}=\frac{b}{h}-\frac{A_{1}^{2}+m_{3}^{2}+\frac{E_{3}^{2}}{h^{2}}}{%
2E_{3}A_{1}}  \label{1c}
\end{equation}

whence%
\begin{equation}
C_{m}-C_{1}=\frac{\left( A_{1}-\sqrt{m_{3}^{2}+\frac{E_{3}^{2}}{h^{2}}}%
\right) ^{2}}{2E_{3}A_{1}}\geq 0\text{.}  \label{cm1}
\end{equation}

After the substitution of (\ref{1c}) back into (\ref{1}), we obtain that%
\begin{equation}
F=\frac{A_{1}^{2}-m_{3}^{2}-\frac{E_{3}^{2}}{h^{2}}}{2A_{1}}.  \label{ae}
\end{equation}

For $E_{1}\geq m_{1}$ it follows from (\ref{a1b}) that 
\begin{equation}
b-\sqrt{b^{2}-1}\leq \frac{A_{1}h}{E_{1}}\leq b-\sqrt{b^{2}-1-h^{2}}\text{.}
\label{ah}
\end{equation}

\section{Conservation of momentum in second order}

If we equate the terms of order $N^{2}$ in (\ref{ez}) taking into account (%
\ref{uz}), (\ref{cz}), (\ref{nz}), (\ref{x24}) we obtain 
\begin{equation}
Y_{L}=Y_{R\text{ }}  \label{LR}
\end{equation}%
where $Y_{L}$ and $Y_{R}$ $\ $\ are corresponding coefficients at $N^{2}$ in 
$Z_{2}-Z_{4}$ and $-Z_{1}-\varepsilon _{3}Z_{3}$, respectively.
Straightforward calculations gives us%
\begin{equation}
Y_{L}\equiv \frac{1}{2\left( X_{2}\right) _{H}}\{m_{4}^{2}-m_{2}^{2}+\frac{%
(E_{1}+E_{2}-E_{3})^{2}-E_{2}^{2}+2\left( X_{2}\right)
_{H}(b_{2}h-1)(E_{1}-E_{3})}{h^{2}}\}+E_{3}(\frac{b}{h}C_{1}-C_{2})
\label{L}
\end{equation}%
and

\begin{equation}
Y_{R}\equiv (\frac{bb_{2}}{h^{2}}-\frac{g_{1}}{2g_{H}h^{2}})\frac{E_{1}^{2}}{%
\sqrt{E_{1}^{2}\frac{(b^{2}-1)}{h^{2}}-m_{1}^{2}}}-\varepsilon _{3}\tau 
\text{,}  \label{R}
\end{equation}%
$\tau $ is given by Eqs. (\ref{ga}) and (\ref{ro}).

In the Kerr case using (\ref{kerr}) - (\ref{rok}) one can see that (\ref{L})
and (\ref{R}) lead to Eq. (4.14) of \cite{j} for $\varepsilon _{3}=-1$.

\section{General bounds on energy and mass}

\subsection{Scenario IN+}

Now, $\varepsilon _{3}=-1$. Then, it follows from (\ref{1}), (\ref{ae}) that%
\begin{equation}
E_{3}^{2}\geq \lambda _{0}^{2},  \label{e3>}
\end{equation}%
\begin{equation}
\lambda _{0}^{2}\equiv h^{2}(A_{1}^{2}-m_{3}^{2})\text{.}  \label{la0}
\end{equation}%
The condition $C_{1}\geq 0$ in (\ref{1c}) gives rise to inequality%
\begin{equation}
E_{3}^{2}-2E_{3}bhA_{1}+h^{2}(A_{1}^{2}+m_{3}^{2})\equiv (E_{3}-\lambda
_{+})(E_{3}-\lambda _{-})\leq 0\text{,}  \label{ea}
\end{equation}%
so%
\begin{equation}
\lambda _{-}\leq E_{3}\leq \lambda _{+}  \label{ein}
\end{equation}%
where 
\begin{equation}
\lambda _{\pm }=h[A_{1}b\pm \sqrt{A_{1}^{2}(b^{2}-1)-m_{3}^{2}}].  \label{la}
\end{equation}

It follows from (\ref{ah}) that the roots satisfy the inequalities 
\begin{equation}
(b-\sqrt{b^{2}-1})b\leq \frac{\lambda _{+}}{E_{1}}\leq \left( b-\sqrt{%
b^{2}-1-h^{2}}\right) (b+\sqrt{b^{2}-1})\text{,}  \label{l+}
\end{equation}%
\begin{equation}
(b-\sqrt{b^{2}-1})^{2}\leq \frac{\lambda _{-}}{E_{1}}\leq (b-\sqrt{%
b^{2}-1-h^{2}})b.  \label{l-}
\end{equation}

As the roots $\lambda _{\pm }$ should be real in the case under discussion, (%
\ref{la}) entails%
\begin{equation}
m_{3}\leq m_{a}\equiv A_{1}\sqrt{b^{2}-1}.  \label{a}
\end{equation}

The condition $m_{3}\leq E_{3}\leq \lambda _{+}$ leads to $m_{3}\leq \lambda
_{+}$ whence 
\begin{equation}
m_{3}\leq m^{\ast }=A_{1}\frac{h}{1+h^{2}}(b+\sqrt{b^{2}-1-h^{2}})\text{.}
\label{mm}
\end{equation}%
Although we have also the bound (\ref{a}), it is easy to show that $m^{\ast
}<m_{a}$. Therefore, the bound (\ref{mm}) is more accurate.

Taking into account (\ref{a1b}) we have%
\begin{equation}
m_{3}\leq m_{B}\equiv \frac{1}{1+h^{2}}(b+\sqrt{b^{2}-1-h^{2}})\left( E_{1}b-%
\sqrt{E_{1}^{2}(b^{2}-1)-m_{1}^{2}h^{2}}\right) .  \label{mb}
\end{equation}%
To have the upper bound for $E_{3}$, we use (\ref{ein}) and the fact that%
\begin{equation}
\lambda _{+}\leq hA_{1}[b+\sqrt{(b^{2}-1)}].
\end{equation}%
Then,%
\begin{equation}
E_{3}\leq E_{B}=(b+\sqrt{b^{2}-1})\left( E_{1}b-\sqrt{%
E_{1}^{2}(b^{2}-1)-m_{1}^{2}h^{2}}\right) \text{.}  \label{eb}
\end{equation}

\subsubsection{Some properties of the bounds}

It is seen from (\ref{mb}) and (\ref{eb}) that%
\begin{equation}
\gamma \equiv \frac{E_{B}}{m_{B}}=\frac{(1+h^{2})(b+\sqrt{b^{2}-1})}{b+\sqrt{%
b^{2}-1-h^{2}}}>1  \label{emb}
\end{equation}%
is a constant.

One can write%
\begin{equation}
\frac{m_{B}}{m_{1}}=f(x)\text{, }x=\frac{E_{1}}{m_{1}}\text{, }f=\frac{1}{%
1+h^{2}}(b+\sqrt{b^{2}-1-h^{2}})\left( xb-\sqrt{x^{2}(b^{2}-1)-h^{2}}\right)
.
\end{equation}%
It is seen that $f(1)=1$. If $E_{1}\gg m_{1}$ or, equivalently, $x\gg 1$,
the function $f\sim x\gg 1$, so $m_{B}\gg m_{1}$. It is monotonic if $h\leq 
\frac{\sqrt{b^{2}-1}}{b}$. If $\frac{\sqrt{b^{2}-1}}{b}<h\leq \sqrt{b^{2}-1}$
this function takes a minimum at 
\begin{equation}
x_{0}=\frac{bh}{\sqrt{b^{2}-1}}\text{, }
\end{equation}%
\begin{equation}
f(x_{0})=\frac{h}{1+h^{2}}\frac{b+\sqrt{b^{2}-1-h^{2}}}{\sqrt{b^{2}-1}}\text{%
.}
\end{equation}%
It is also instructive to look at the ratio $\frac{m_{B}}{E_{1}}=\chi (x)$
where%
\begin{equation}
\chi =\frac{1}{1+h^{2}}(b+\sqrt{b^{2}-1-h^{2}})\left( b-\sqrt{b^{2}-1-\frac{%
h^{2}}{x}}\right)  \label{hi}
\end{equation}%
\ it is seen that $\frac{d\chi }{dx}<0$. Here, $\chi (1)=1$ Therefore, $\chi
<1$ for $x>1$, so $m_{B}<E_{1}.$ Therefore, the ratio $\frac{E_{B}}{E_{1}}$
attains the maximum value for $E_{1}=m\,_{1}$. Then, $\frac{E_{B}}{E_{1}}$
is given by Eq. (\ref{emb}). It is seen from (\ref{la}) and (\ref{eb}) that $%
\lambda _{+}=E_{B}$ if and only if $m_{3}=0$.

Let $E_{3}=\lambda _{+}$. If $m_{3}\geq A_{1}$, Eq. (\ref{e3>}) is satisfied
automatically. Let $m_{3}\leq A_{1}$, $\sqrt{b^{2}-1}A_{1}$. It is seen from
(\ref{la}) and (\ref{e3>}) that%
\begin{equation}
\lambda _{+}\geq hA_{1}b\geq \lambda _{0}\text{.}  \label{+0}
\end{equation}

If $E_{3}=\lambda _{\pm }$, it follows from (\ref{1c}), (\ref{ea}) that $%
C_{1}=0$.

Now, it is instructive to consider some consequences of Eqs. (\ref{a1b}), (%
\ref{la}) for massive and massless particles 1 and 3.

Particles 1 and 3 are massless: 
\begin{equation}
\lambda _{+}=E_{1}.  \label{l=e}
\end{equation}

Particle 1 is massless, particle 3 $\ $is massive.\ Then,%
\begin{equation}
A_{1}=E_{1}(\frac{b-\sqrt{b^{2}-1}}{h})\text{, }\lambda _{+}<hA_{1}(b+\sqrt{%
b^{2}-1})\text{,}
\end{equation}%
so%
\begin{equation}
\lambda _{+}<E_{1}.
\end{equation}%
As in the limit $m_{1}\ll E_{1}$, we have $\lambda _{+}\leq E_{1},$ there is
no net energy extraction in this case.

Particle 1 is massive, particle 3 is massless. Then, 
\begin{equation}
A_{1}>E_{1}(\frac{b-\sqrt{b^{2}-1}}{h})\text{, }\lambda _{+}=hA_{1}(b+\sqrt{%
b^{2}-1})\text{,}
\end{equation}%
so%
\begin{equation}
\lambda _{+}>E_{1}\text{. }
\end{equation}

For the particular case described by Eq. (\ref{bbh}), all aforementioned
properties agree with those for the Kerr black hole \cite{j}.

\subsection{Scenario OUT$+$}

In this scenario, the condition $C_{1}\geq 0$ should be satisfied as well as
in case IN$+$ considered before.\ Therefore, Eq. (\ref{ein}) holds.

Eqs. (\ref{1}), (\ref{ae}) and the condition $\varepsilon _{3}=+1$ give us 
\begin{equation}
E_{3}^{2}\leq \lambda _{0}^{2}\text{, }  \label{el<}
\end{equation}%
\begin{equation}
m_{3}\leq A_{1}  \label{ma}
\end{equation}%
instead of (\ref{e3>}). Thus we have two upper bounds $E_{3}\leq \lambda
_{0} $ and $E_{3}\leq \lambda _{+}$. Because of the property (\ref{+0}), the
bound (\ref{el<}) is more relevant, so%
\begin{equation}
\lambda _{-}\leq E_{3}\leq \lambda _{0}.  \label{-e0}
\end{equation}

It is seen from (\ref{ah}) that 
\begin{equation}
\lambda _{0}\leq hA_{1}\leq \lambda _{1}\equiv E_{1}(b-\sqrt{b^{2}-1-h^{2}}).
\label{01}
\end{equation}%
If $\frac{\lambda _{1}}{E_{1}}<1$, there is no net energy extraction since
in this case $E_{3}<E_{1}$. In particular, this happens in Kerr case when $%
\frac{\lambda _{1}}{E_{1}}=2-\sqrt{2}$ and this is the reason why this
scenario was rejected in \cite{j}. However, for a more general metric, we
may try to obtain $\lambda _{1}>E_{1}$ that gives the necessary condition%
\begin{equation}
\sqrt{1+h^{2}}<b<1+\frac{h^{2}}{2}\text{.}  \label{1bh}
\end{equation}

Consistency of Eqs. (\ref{ein}) and (\ref{el<}) requires%
\begin{equation}
\lambda _{-}\leq \lambda _{0}\text{.}
\end{equation}%
Then one can find from (\ref{l-}) that%
\begin{equation}
(b-\sqrt{b^{2}-1})^{2}\leq b-\sqrt{b^{2}-1-h^{2}}\text{.}  \label{b1h}
\end{equation}%
One can observe that $(b-\sqrt{b^{2}-1})^{2}\leq b-\sqrt{b^{2}-1}\leq b-%
\sqrt{b^{2}-1-h^{2}}$, so (\ref{b1h}) is satisfied. Thus the inequality $%
\lambda _{1}>E_{1}$ is indeed possible, so the case OUT$+$ is of some
potential interest for the energy extraction in contrast to the Kerr case 
\cite{p}, \cite{j},

\subsection{Scenario OUT$-$}

As $\varepsilon _{3}=+1$, inequalities (\ref{el<}) and (\ref{ma}) should be
still satisfied since they follow from $F\geq 0$, where $F$ is given by (\ref%
{ae}). Eq. (\ref{01}) that follows from $E_{3}\leq \lambda _{0}$ is valid as
well. As we want to have the possibility of the energy extraction we must
assume (\ref{1bh}).

This is not the end of story. To check the possibility of such a variant, we
must also take into account the condition (\ref{cneg}) where $C_{1}$ is
given by Eq. (\ref{1c}) and verify its compatibility with (\ref{el<}) and (%
\ref{1bh}). Then, from (\ref{1c}) we have that%
\begin{equation}
\frac{b}{h}<\frac{A_{1}^{2}+m_{3}^{2}+\frac{E_{3}^{2}}{h^{2}}}{2E_{3}A_{1}},
\end{equation}%
so that%
\begin{equation}
(E_{3}-\lambda _{+})(E_{3}-\lambda _{-})>0.  \label{e+}
\end{equation}

Now, we will consider different possibilities separately.

a) The roots (\ref{la}) $\lambda _{+}$ and $\lambda _{-}$ of (\ref{e+}) are
complex, 
\begin{equation}
A_{1}^{2}(b^{2}-1)<m_{3}^{2}\leq A_{1}^{2},b^{2}<2.  \label{mb2}
\end{equation}%
As a result, there is no bound on $\frac{E_{3}}{E_{1}}$ from (\ref{e+}).
However, the bound (\ref{el<}) persists. If (\ref{1bh}) is satisfied, then
the upper limit $\lambda _{1}>E_{1}$. In addition, it follows from (\ref{1bh}%
) \ and (\ref{mb2}) that $h<1.$

Let now the roots $\lambda _{+}$ and $\lambda _{-}$ $\ $be real, so 
\begin{equation}
m_{3}^{2}\leq A_{1}^{2}(b^{2}-1).  \label{mab}
\end{equation}

b) $E_{3}>\lambda _{+}$.

This case is inconsistent with (\ref{el<}) and (\ref{+0}) and should be
rejected.

c) $E_{3}<\lambda _{-}.$ If $\frac{\lambda _{-}}{E_{1}}<1$ and $\frac{%
\lambda _{0}}{E_{1}}<1$ , no energy extraction is possible. To avoid this
case of no interest, we require that $\lambda _{-}>E_{1}$ and $\lambda
_{0}>E_{1}$. The first condition requires (\ref{1bh}). The second one leads
to the bound for the mass:%
\begin{equation}
m_{3}^{2}\leq \frac{E_{1}^{2}}{h^{2}}[\left( b-\sqrt{b^{2}-1-h^{2}}\right)
^{2}-1]  \label{m3}
\end{equation}%
where we took into account (\ref{ah}). The right hand side of (\ref{m3}) is
nonnegative if (\ref{1bh}) is satisfied.

\section{Energy extraction and unconditional upper limits on its efficiency}

It follows from (\ref{LR}) - (\ref{R}) that 
\begin{equation}
m_{4}^{2}+2\left( X_{2}\right) _{H}S=m_{2}^{2}+\frac{%
2E_{2}(E_{3}-E_{1})-(E_{1}-E_{3})^{2}+2\left( X_{2}\right)
_{H}(b_{2}h-1)(E_{3}-E_{1})}{h^{2}},  \label{m4}
\end{equation}%
\begin{equation}
S=-Y_{R}+(\frac{b_{2}}{h}C_{1}-C_{2})E_{3}.  \label{s}
\end{equation}

If $S\geq 0$ (the corresponding conditions are discussed in Appendix B), the
left hand side of (\ref{m4}) is positive (or, at lest, nonnegative) and this
gives us the constraint on $E_{2}$. Since the case $E_{1}\geq E_{3}$ is not
interesting, we assume that $E_{1}<E_{3}$. Then, if follows from (\ref{m4})
that%
\begin{equation}
E_{2}\geq \frac{1}{2}[(E_{3}-E_{1})-\frac{m_{2}^{2}h^{2}}{E_{3}-E_{1}}]-\nu
\equiv \kappa \text{, }\nu \equiv \left( X_{2}\right) _{H}(b_{2}h-1).
\label{e2}
\end{equation}%
In the Kerr case, $\nu =0$, $h=1$ and (\ref{e2}) reduces to Eq. (4.15) of 
\cite{j}.

We want to derive the unconditional upper limit on $\frac{E_{3}}{E_{1}}$
using this inequality, when needed. Below, we discuss different scenarios
separately.

\subsection{Scenario IN+}

To gain the highest efficiency of extraction, we want to have $E_{3}=\lambda
_{+}$ which is the maximum possible value according to (\ref{ein}). In this
case, according to (\ref{1c}) and (\ref{ea}), $C_{1}=0$, \ so $\delta \geq 0$
gives us $C_{2}\geq 0$. To maximize the possible outcome, we concentrate on
the case when $\lambda _{+}=E_{B}$ given by Eq. (\ref{eb}). In turn, this
implies that $m_{3}=0$.

The efficiency of the possible energy extraction is given by the quantity%
\begin{equation}
\eta =\frac{E_{3}}{E_{1}+E_{2}}.  \label{nu}
\end{equation}

As usual, we assume that $E_{2}\geq m_{2}$, so a particle is injected from
infinity. As we have two conditions $E_{2}\geq m_{2}$ and $E_{2}\geq \kappa $
we must consider two relationships between $m_{2}$ and $\kappa $.

a) $\kappa >m_{2}$. Then, it follows from (\ref{e2}) that $m_{2}<m_{+}$ where%
\begin{equation}
m_{+}=\frac{y}{h^{2}}(\sqrt{1+h^{2}-2\frac{\nu h^{2}}{y}}-1)\,,\text{ }%
y=\lambda _{+}-E_{1}>0\text{.}  \label{m+}
\end{equation}

For the efficiency we obtain%
\begin{equation}
\eta \leq \eta _{m}=\frac{\lambda _{+}}{E_{1}+\kappa }\text{.}  \label{ak1}
\end{equation}%
We are interested in the cases when extraction of energy is possible, so $%
\eta _{m}>1$. Using (\ref{e2}) and the fact that now $E_{3}=\lambda _{+}$,
we obtain 
\begin{equation}
y^{2}+2\nu y+m_{2}^{2}h^{2}>0\text{.}  \label{nl}
\end{equation}

If $b_{2}h-1\geq 0$, so $\nu \geq 0$, then inequality (\ref{nl}) is
satisfied automatically. In this case, the positivity of the expression
inside the radical in (\ref{m+}) implies also%
\begin{equation}
\frac{2\nu h^{2}}{1+h^{2}}<E_{1}(\gamma -1)  \label{e1n}
\end{equation}%
where $\gamma $ is given by (\ref{emb}).

If $\nu <0$, there are different options for Eq. (\ref{nl}) to be satisfied.
(i) $m_{2}h>\left\vert \nu \right\vert $, then (\ref{nl}) is satisfied for
any $y$, so $\eta _{m}>1$ always. In two other cases additional constrains
are required: (ii) $y\leq y_{-}$, so $\lambda _{+}\leq E_{1}+y_{-}$, (iii) $%
y\geq y_{+}$, so $\lambda _{+}\geq E_{1}+y_{+}$ where $y_{\pm }=\left\vert
\nu \right\vert \pm \sqrt{\left\vert \nu \right\vert ^{2}-m_{2}^{2}h^{2}}$
are roots of (\ref{nl}).

b) $\kappa \leq m_{2}$. Then, $m_{2}\geq m_{+}$, 
\begin{equation}
\eta \leq \eta _{m}=\frac{\lambda _{+}}{E_{1}+m_{2}}\text{.}  \label{k<m}
\end{equation}%
The condition $\eta _{m}>1$ requires $m_{2}<\lambda _{+}-E_{2}$.

Now, we want to find some general unconditional upper bound on $\eta $. We
want to maximize $\eta _{m}$, thus minimizing $E_{2}$ as a function of $%
m_{2} $ for given $E_{1}$, $m_{1}$. As we must have simultaneously $%
E_{2}\geq m_{2} $ (particle 2 is injected from infinity) and $E_{2}\geq
\kappa $, this corresponds to $m_{2}=\kappa $, whence $m_{2}=m_{+}$ with $%
m_{+}$ given by (\ref{m+}). To get the possible maximum $\eta $, we also put 
$E_{1}=m_{1}$ and $\lambda _{+}=E_{B}$.

As \ a result, we have%
\begin{equation}
\eta _{m}=\frac{\lambda _{+}}{m_{1}+\kappa }.
\end{equation}%
Now, taking into account (\ref{eb}), (\ref{m+}) we can write%
\begin{equation}
\lambda _{+}=qm_{1}\text{, }q=(b+\sqrt{b^{2}-1})(b-\sqrt{b^{2}-1-h^{2}}),
\end{equation}%
\begin{equation}
1\leq q\leq 1+h^{2}\leq b^{2}\text{.}  \label{q}
\end{equation}%
\begin{equation}
m_{+}=sm_{1}\text{, }s=\frac{1-q+\sqrt{(1+h^{2})(q-1)^{2}-2\nu h^{2}(q-1)}}{%
h^{2}}
\end{equation}%
\begin{equation}
\eta _{m}=\frac{\lambda _{+}}{m_{1}+m_{2}}=\frac{qh^{2}}{1+h^{2}-q+\sqrt{%
(1+h^{2})(q-1)^{2}-2\nu (q-1)}}
\end{equation}%
These formulas are simplified when $\nu =0$. Then,%
\begin{equation}
s=\frac{q-1}{h^{2}}\left( \sqrt{1+h^{2}}-1\right) \text{,}
\end{equation}%
\begin{equation}
\eta _{m}=\frac{q(\sqrt{1+h^{2}}+1)}{q+\sqrt{1+h^{2}}}>1.  \label{lim}
\end{equation}%
It is also seen from (\ref{q}), (\ref{lim}) that $\eta \,_{m}\leq q\leq
1+h^{2}\leq b^{2}$. When $b\gg 1+h^{2}$, one obtains that $q\approx 1+h^{2}$
and $\eta _{m}\approx \sqrt{1+h^{2}}$.

In the Kerr case, $\nu =0$, $q=(2+\sqrt{3})(2-\sqrt{2})$, $s=(\sqrt{2}%
-1)(q-1),$%
\begin{equation}
\eta _{m}=\frac{2(2+\sqrt{3})}{q+2}\approx 1.466
\end{equation}%
that agrees with the results of \cite{p}, \cite{j}.

\subsection{Scenario OUT +}

Now, according to (\ref{+0}), (\ref{el<}), 
\begin{equation}
\eta _{m}=\frac{\lambda _{0}}{E_{1}+E_{2}}  \label{out}
\end{equation}%
where we put $E_{3}=\lambda _{0}$ instead of $E_{3}=\lambda _{+}$ typical of
IN+ case. As usual, we choose $E_{1}=m_{1}$ as a reference point. Apart from
this, now we do not have the condition $C_{1}=0$ that followed from $%
E_{3}=\lambda _{+}$ in the aforementioned scenario. Correspondingly, there
is no definite restriction on $S$ in (\ref{m4}) and there is no bound on $%
E_{2}$ similar to (\ref{e2}). Therefore, we put $E_{2}=m_{2}$. To gain the
maximum efficiency, we simply choose $m_{2}=0$. Then,%
\begin{equation}
\eta \leq \frac{\lambda _{0}}{m_{1}}\leq \eta _{m}=b-\sqrt{b^{2}-1-h^{2}},
\label{l0}
\end{equation}%
where (\ref{ae}), (\ref{la0}) where taken into account. Eq. (\ref{l0})
corresponds to $m_{3}=0$ and gives the unconditional limit for the scenario
under discussion. It is possible to gain $\eta _{m}>1$, provided Eq. (\ref%
{1bh}) is satisfied. This option is absent for the Kerr case $b=2$, $h=1$.
For a given $h$, the maximum of $\eta _{m}$ is achieved at $b=\sqrt{1+h^{2}%
\text{ }}$when $\eta _{m}=\sqrt{1+h^{2}}>1$.

\subsection{Scenario OUT $^{\_}$}

Now, we must enumerate all cases already considered in Sec. VIII\ C and
apply the corresponding results to the evaluation of the extraction
efficiency. In doing so, we put $E_{1}=m_{1}$, $m_{2}=0$.

a) Eq. (\ref{mb2}) is valid. There is no bound from Eq. (\ref{e+}), so the
only bound is $E_{3}\leq \lambda _{0}$. Then, Eq. (\ref{l0}) applies as well
as subsequent discussion, so $\eta _{m}>1$ is possible. Consistency of (\ref%
{mb2}) and (\ref{1bh}) requires also $h<1$.

b) Eq. (\ref{mab}) is valid. Then, there is competition between two bounds $%
E_{3}\leq \lambda _{0}$ and $E_{3}\leq \lambda _{-}$. As $\lambda _{0}$ (\ref%
{la0}) is a monotonically decreasing function of $m_{3}$ and $\lambda _{-}$
is a monotonically increasing one, $E_{3}\leq \lambda ^{\ast }$ where $%
\lambda ^{\ast }$ corresponds to the point of their intersection. In that
point,%
\begin{equation}
m_{3}^{2}=A_{1}^{2}\frac{b^{2}-1}{b^{2}}\text{,}  \label{m3a}
\end{equation}%
\begin{equation}
\eta _{m}=\frac{\lambda ^{\ast }}{m_{1}}=\frac{b-\sqrt{b^{2}-1-h^{2}}}{b}<1%
\text{,}
\end{equation}%
so there is no energy extraction.

We would like to remind that in IN+ scenario, the massless case $m_{3}=0$
was favorite. This is not so in the scenario under discussion since it is
the condition (\ref{mb2}) that is consistent with $\eta _{m}>1$ but this
implies that $m_{3}\neq 0$.

\section{Examples of reactions}

In addition to general bounds, we illustrate the efficiency of energy
extraction by some simple examples. For simplicity, in all scenarios we take 
$\nu =0$.

\subsection{Scenario IN+}

\subsubsection{Elastic collision}

For simplicity, we choose $m_{1}=m_{2}=m_{3}=m_{4}=m_{0}$, $E_{1}=m_{0}$.

Then, it follows from (\ref{a1b}), (\ref{la}) that%
\begin{equation}
\frac{\lambda _{+}}{m_{0}}\equiv \mu =b(b-\sqrt{b^{2}-1-h^{2}})+\sqrt{Y^{2}}%
\text{, }Y^{2}=(b^{2}-1)(b-\sqrt{b^{2}-1-h^{2}})^{2}-h^{2}.
\end{equation}%
If%
\begin{equation}
\frac{\sqrt{b^{2}-1}}{b}\leq h\leq \sqrt{b^{2}-1}\text{,}  \label{hint}
\end{equation}%
it turns out that%
\begin{equation}
\left\vert Y\right\vert =b^{2}-1-\sqrt{b^{2}-1-h^{2}}b,  \label{y}
\end{equation}%
so

\begin{equation}
\mu =2b^{2}-1-2b\sqrt{b^{2}-1-h^{2}}>1.  \label{lm}
\end{equation}%
One can see that%
\begin{equation}
1<\mu <2b^{2}-1,  \label{m1}
\end{equation}%
where the minimum and maximum values correspond to (\ref{hint}). For $b\gg
1+h^{2}$, 
\begin{equation}
\mu \approx 1+h^{2}\text{.}  \label{mlar}
\end{equation}

For the Kerr case, it follows from (\ref{bbh}) that%
\begin{equation}
\mu =7-4\sqrt{2}\approx 1.343  \label{mu}
\end{equation}%
in agreement with Eq. (5.1) of \cite{j}.

To evaluate the efficiency of the extraction process $\eta =\frac{E_{3}}{%
E_{1}+E_{2}}$, we take $E_{3}=\lambda _{+}.$ From (\ref{e2}), (\ref{m+}) we
have now%
\begin{equation}
\frac{\kappa }{m_{0}}=\frac{\mu ^{2}-2\mu +1-h^{2}}{2(\mu -1)}\text{,}
\label{k0}
\end{equation}%
\begin{equation}
\frac{m_{+}}{m_{0}}=\frac{\mu -1}{1+\sqrt{1+h^{2}}}\text{.}  \label{mm0}
\end{equation}

a) $\kappa >m_{2}$. Now, according to (\ref{m+}), $m_{2}<m_{+}$ where now it
follows from (\ref{mm0}) that we must have%
\begin{equation}
\mu -1>1+\sqrt{1+h^{2}}.  \label{m11}
\end{equation}

The maximum value of the left hand side equals $2(b^{2}-1)$, so the
necessary condition is $b^{2}>\frac{3}{2}+\frac{\sqrt{1+h^{2}}}{2}$.

Then, the upper limit is given by (\ref{ak1}) where $\kappa $ is taken from (%
\ref{k0}),%
\begin{equation}
\eta _{m}=\frac{2\mu (\mu -1)}{\mu ^{2}-1-h^{2}}\text{.}  \label{nmh}
\end{equation}

In particular, it follows from (\ref{mlar}), (\ref{nmh}) that for large $b$, 
$\eta _{m}\approx 2$.

For the Kerr metric, inequality (\ref{m11}) is not satisfied, so this case
is absent.

b) $\kappa \leq m_{2}$, $m_{2}\geq m_{+}$. Then, Eq. (\ref{k<m}) applies,%
\begin{equation}
\eta _{m}=\frac{\mu }{2}\text{.}  \label{mu2}
\end{equation}

For the Kerr case $\eta _{m}\approx 0.672<1$, so there is no energy
extraction in agreement with Eq. (5.2) of \cite{j}. However, if $b$ is
sufficiently large, it is possible to have $\eta _{m}>1$. \ Indeed, for $%
b\gg 1+h^{2}$ one sees from (\ref{mlar}) that $\eta _{m}\approx \frac{1+h^{2}%
}{2}$, so for $h>1$ we obtain that $\eta _{m}>1$.

If, instead of (\ref{hint}), 
\begin{equation}
h<\frac{\sqrt{b^{2}-1}}{b}\text{,}
\end{equation}%
the quantity $Y$ in (\ref{y}) is such that%
\begin{equation}
\left\vert Y\right\vert =\sqrt{b^{2}-1-h^{2}}b+1-b^{2}\text{,}
\end{equation}%
\begin{equation}
\frac{\lambda _{+}}{m_{0}}=1\text{,}
\end{equation}

and there is no energy extraction.

Let now $m_{1}=m_{3}\equiv m_{0}$, $m_{2}=m_{4}$ but $m_{2}\neq m_{0}$. We
can optimize $\eta $ by taking $m_{2}=\kappa =m_{+}$ where $m_{+}$ is given
by (\ref{mm0}). Then, $\eta _{m}$ is given by (\ref{ak1}), 
\begin{equation}
\eta _{m}=\frac{\mu (1+\sqrt{1+h^{2}})}{\sqrt{1+h^{2}}+\mu }>1\text{.}
\label{nm1}
\end{equation}

For large $\mu $, $\eta _{m}\approx 1+\sqrt{1+h^{2}}$.

In the Kerr case (\ref{mu}) Eq. (\ref{nm1}) reduces to%
\begin{equation}
\eta _{m}=\frac{18\sqrt{2}+11}{13}\text{,}
\end{equation}%
so we return to Eq. (5.4) of \cite{j}.

\subsubsection{Compton scattering}

To gain the maximum of efficiency, we take $m_{3}=0$ since it is the
condition of having $\lambda _{+}=E_{3}=E_{B}$. This is explained in
discussion after Eq. (\ref{hi}) that generalize the corresponding
observations in \cite{p}, \cite{j}.

1) $m_{1}=m_{3}=0$, $m_{2}=m_{4}=m_{0}$

Then, it follows from (\ref{l=e}) that $\lambda _{+}=E_{1\text{ }},$so $%
C_{1}=0$ according to (\ref{1c}). Eqs. (\ref{LR}) - (\ref{R}) with $C_{1}=0$
give us that $C_{2}=0$. It means that in the given approximation, particle 3
is critical and one cannot distinguish between particles 1 and 3 at all.
Actually, one cannot detect scattering in this approximation, so the
situation is similar to that in the Kerr case \cite{j}.

2) $m_{4}=m_{1}=m_{0}$, $m_{2}=m_{3}=0$, $E_{1}=m_{0}$

It follows from (\ref{la}) that%
\begin{equation}
\mu =\frac{\lambda _{+}}{m_{0}}=(b+\sqrt{b^{2}-1})(b-\sqrt{b^{2}-1-h^{2}})>1.
\label{lm1}
\end{equation}

Now, $m_{2}=0<\kappa =\frac{\mu -1}{2}$. \ According to (\ref{ak1}),%
\begin{equation}
\eta _{m}=\frac{2\mu }{\mu +1}
\end{equation}%
For sufficiently large $b$ and, hence, large $\mu $ the upper limit can
approach 2. In the Kerr case, 
\begin{equation}
\mu =\left( 2+\sqrt{3}\right) (2-\sqrt{2})\approx 2.186  \label{mkerr}
\end{equation}%
and%
\begin{equation}
\eta _{m}=\frac{2(54+14\sqrt{3}-10\sqrt{2}+\sqrt{6})}{97}
\end{equation}%
in agreement with Eq. (5.7) of \cite{j}.

\subsubsection{Pair annihilation}

Let two massive particles collide to produce two massless ones,%
\begin{equation}
m_{1}=m_{2}=m_{0}\text{, }m_{3}=m_{4}=0\text{.}
\end{equation}%
For definiteness,%
\begin{equation}
E_{1}=m_{0}\text{,}
\end{equation}%
as usual. Then, we obtain again that (\ref{lm1}), (\ref{mm0}) and (\ref{k0})
hold. Therefore, for a) $\kappa >m_{0}$ Eq. (\ref{nmh}) is valid, for b) $%
\kappa \leq m_{0}$, Eq. (\ref{mu2}) applies as well as corresponding
discussion. For the Kerr metric, only case b) is realized with (\ref{mkerr})
and $\eta _{m}=\frac{\mu }{2}\approx 1.093$, so we return to Eqs. (5.8),
(6.9) of \cite{j}$.$

\subsection{Scenario OUT+}

\subsubsection{Elastic collision}

Let $m_{1}=m_{2}=m_{3}=m_{4}=m_{0}$, $E_{1}=m_{0}$. Then, it follows from (%
\ref{out}) that

\begin{equation}
\eta _{m}=\frac{\lambda _{0}}{2m_{0}}\text{.}  \label{la2}
\end{equation}%
According to (\ref{a1b}), (\ref{e3>}),%
\begin{equation}
\eta _{m}=\frac{\left( b-\sqrt{b^{2}-1-h^{2}}\right) ^{2}-h^{2}}{2}<1,
\label{el}
\end{equation}%
so there is no energy extraction.

\subsubsection{Compton scattering}

1) $m_{1}=m_{3}=0$, $m_{2}=m_{4}=m_{0}=E_{1}$

Now, $\eta _{m}$ is given by $\eta _{m}=\frac{\lambda _{0}}{m_{0}}$. It
follows from (\ref{a1b}), (\ref{e3>}) that%
\begin{equation}
\lambda _{0}=hA_{1}=m_{0}(b-\sqrt{b^{2}-1})\text{,}
\end{equation}%
\begin{equation}
\eta _{m}=b-\sqrt{b^{2}-1}<1\text{,}
\end{equation}%
so there is no net energy extraction.

2) $m_{4}=m_{1}=m_{0}$, $m_{2}=m_{3}=0$, $E_{1}=m_{0}$

In a similar manner,%
\begin{equation}
\lambda _{0}=hA_{1}=m_{0}(b-\sqrt{b^{2}-1-h^{2}})\text{,}  \label{laha}
\end{equation}%
so

\begin{equation}
\eta _{m}=b-\sqrt{b^{2}-1-h^{2}}\text{.}  \label{nul}
\end{equation}%
One can check that $\eta _{m}>1$ if (\ref{1bh}) holds.

In particular, for $b=\sqrt{1+h^{2}}$, $\eta _{m}=\sqrt{1+h^{2}}$\thinspace $%
>1$.

\subsubsection{Pair annihilation}

We choose for simplicity

\begin{equation}
m_{1}=m_{2}=m_{0}=E_{1}\text{, }m_{3}=m_{4}=0.
\end{equation}%
Now, Eq. (\ref{laha}) applies,%
\begin{equation}
\eta _{m}=\frac{b-\sqrt{b^{2}-1-h^{2}}}{2}\text{.}
\end{equation}

It is seen that $\eta _{m}>1$ if $b^{2}<\frac{5+h^{2}}{4}$.

The cases with $m_{4}=0$ are of no interest since one can check that they
give no energy extraction.

\subsection{Scenario OUT $-$}

According to previous analysis, there are no new interesting options here
with $\eta _{m}>1$, so the situation is similar to the previous case.

\section{Summary and conclusion}

We investigated different types of high-energetic processes near black
holes. As the BSW effect means that the energy in the centre of mass frame $%
E_{c.m.}$ of two colliding particles grows unbound, it would seem that this
effect is favorite for observation of high-energy particles at infinity.
These intuitive expectations are not confirmed. It turned out that there are
bounds on the mass and energy of particles created due to the BSW effect. It
is possible in some cases to gain more energy than it was injected but the
excess of energy at infinity is restricted. In doing so, we generalized the
corresponding properties of the Kerr metric \cite{p}, \cite{j}.

Meanwhile, there are some new features which exist for dirty black holes
only and are absent in the Kerr case. First of all, it concerns the possible
choice of scenario of collisions. In the Kerr case, the only potential
scenario that gives energy extraction is (in our notations) IN+ when a
particle immediately after collision moves inwardly, bounces and only
afterwards goes to infinity \cite{ps}, \cite{p}, \cite{j}. Meanwhile, it
turned out that in general two other scenarios OUT $+$ and OUT $-$ are also
possible, so a particle can escape to infinity directly. The value of energy
extraction depends crucially on relationship between just two numbers ($b$
and $h$) that characterize near-horizon behavior of the metric. Account for
generic $b$ and $h$ not only opens new scenarios for energy extraction but
also extends significantly the diversity of possible cases within the old
scenario IN$+$ typical of the Kerr metric. For $b\sim h\sim 1$, the maximum
possible extraction efficiency $\eta _{m}\,\ \,$may be higher than in the
Kerr case being enhanced by numeric factors like 2 or even higher. In some
cases, $\eta _{m}$ even grows formally without limit \ but this also
requires "exotic" situations with large $b$ or $h$, so in effect $\eta _{m}$
remains limited. Thus in spite of essential extension of the whole picture,
the main conclusion does not change radically from the Kerr case. In
principle, the corresponding bounds on the energy become weaker but the
bounds persist. The results for the unconditional upper limit for different
scenarios are summarized in Table 1.

\bigskip

\begin{tabular}{|l|l|l|}
\hline
Scenario & $\eta _{m}>1$ for unconditional upper limit & Upper bound on $%
E_{3}$ \\ \hline
IN+ & always (Kerr black hole included) & $\lambda _{+}$ \\ \hline
OUT+ & $\sqrt{1+h^{2}}<b<1+\frac{h^{2}}{2}$ & $\lambda _{0}$ \\ \hline
OUT- & $\sqrt{1+h^{2}}<b<1+\frac{h^{2}}{2}$, $b<\sqrt{2}$, $h<1$ & $\lambda
_{0}$ \\ \hline
\end{tabular}

\bigskip Table 1: Scenarios with possible extraction of energy and their
main properties.

It is seen that scenario IN+ is the most favorable, scenario OUT- is the
most restrictive.

One can say that the BSW effect and the Penrose process due to near-horizon
collisions are mutually "unfriendly". Thus there exist serious difficulties
for direct observation of the consequences of the BSW effect although they
become milder if one deals with a dirty black hole instead of the Kerr
metric. However, this effect can leave indirect imprint at infinity, since
new channels of reactions can be open which are forbidden in the laboratory
experiments.

The analysis of elementary acts of collision from the first principles
suggested in the present work, can be used also for the analysis of
high-energy processes in the vicinity of charged static black holes. This
will be done elsewhere.

\begin{acknowledgments}
I thank Igor Tanatarov for reading the manuscript and useful comments. This
work was supported in part by the Cosmomicrophysics section of the Programme
of the Space Research of the National Academy of Sciences of Ukraine.
\end{acknowledgments}

\section{Appendix A: Special case: near-horizon circular orbit}

This appendix generalizes the contents of Appendix A in \cite{j}. Formally,
in Eq. (\ref{R}) divergences appear if $E_{1}^{2}=E_{0}^{2}\equiv \frac{%
m^{2}h^{2}}{b^{2}-1}$. Actually, this case corresponds to a "circular" orbit
with the constant proper distance ($Z_{1}=0$) near the horizon and requires
some inessential changes. The circular orbit is defined by equations

\begin{equation}
Z^{2}=(E-\omega L)^{2}-N^{2}(m^{2}+\frac{L^{2}}{g})=0  \label{v}
\end{equation}%
and%
\begin{equation}
\frac{dZ^{2}}{dN}=0.  \label{v1}
\end{equation}%
It follows from (\ref{v}), (\ref{v1}) that%
\begin{equation}
E-\omega L=N\sqrt{m^{2}+\frac{L^{2}}{g}}  \label{en}
\end{equation}%
and%
\begin{equation}
-\sqrt{m^{2}+\frac{L^{2}}{g}}L\frac{d\omega }{dN}-(m^{2}+\frac{L^{2}}{g})+N%
\frac{L^{2}}{2g^{2}}\frac{dg}{dN}=0  \label{ln}
\end{equation}

Near the horizon where $N\ll 1$ one can find from (\ref{en}) and (\ref{ln})
that%
\begin{equation}
E=E_{0}(1+e_{1}N+e_{2}N^{2})+O(N^{3})\text{,}  \label{e1}
\end{equation}%
\begin{equation}
L=\frac{E_{0}}{\omega _{H}}(1+\gamma _{1}N+\gamma _{2}N^{2})+O(N^{2}).
\end{equation}%
Then, one can obtain from (\ref{om}) and (\ref{ln}) that%
\begin{equation}
\gamma _{1}=e_{1}=\frac{2b_{2}b-\frac{g_{1}}{g_{H}}}{b^{2}-1}\text{,}
\end{equation}%
\begin{equation}
\frac{\omega _{H}L}{E}=1+\tilde{\gamma}N^{2}+O(N^{3}).
\end{equation}%
\begin{equation}
\tilde{\gamma}=\gamma _{2}-e_{2}\text{,}
\end{equation}%
\begin{equation}
X=\frac{b}{h}N+E_{0}N^{2}(\frac{b}{h}e_{1}-\tilde{\gamma}-\frac{b_{2}}{h}%
)+O(N^{3}).
\end{equation}

We do not list the coefficients $e_{2}$, $\gamma _{2}$ explicitly since they
are too cumbersome. For our purpose, their exact values are irrelevant, it
is sufficient to know that they are finite. Then, Eq. (\ref{1}) obtained
from the terms of the order $N$ does not change since one may put there $%
E=E_{0}$ directly. Eq. (\ref{LR}) somewhat changes since now the momentum
conservation in the second order reads $\tilde{Y}_{L}=\tilde{Y}_{R}$ where 
\begin{equation}
\tilde{Y}_{L}=E_{0}(\frac{b}{h}e_{1}-\tilde{\gamma})+Y_{L}  \label{y1}
\end{equation}%
and

\begin{equation}
\tilde{Y}_{R}\equiv -\varepsilon _{3}\tau \text{.}  \label{r1}
\end{equation}%
Here, $Y_{L}$ is obtained from (\ref{L}) by substitution $E_{1}\rightarrow
E_{0}$. What is important is that the quantity $m_{4}$ is still finite. In
the Kerr case $E_{0}=\frac{m}{\sqrt{3}}$, $e_{1}=4/3$, the coefficient $%
\tilde{\gamma}=1$, so Eq. (\ref{e1}) corresponds to Eq. (A1) and Eq. $\tilde{%
Y}_{L}=\tilde{Y}_{R}$ with (\ref{y1}), (\ref{r1}) corresponds to Eq. (A4) of 
\cite{j}.

\section{Appendix B: condition $S\geq 0$}

The contents of the present Appendix generalizes Appendix B of \cite{j}. Let
us consider the quantity $Y_{R}$ (\ref{R}) which represents the right hand
side of Eq. (\ref{LR}) that follows from the momentum conservation (\ref{ez}%
) in the second order with respect to $N$. We are interested in the case $%
C_{1}=0$ when $E_{3}=\lambda _{+}$ and (\ref{s}) reduces to%
\begin{equation}
S=S_{1}+S_{2}\text{, }
\end{equation}%
\begin{equation}
S_{1}=\frac{1}{h^{2}}(bb_{2}-\frac{g_{1}}{2g_{H}})[-\varepsilon _{3}\frac{%
\lambda _{+}^{2}}{\sqrt{\lambda _{+}^{2}[\frac{b}{h})^{2}-\frac{1}{h^{2}}%
]-m_{3}^{2})}}-\frac{E_{1}^{2}}{\sqrt{E_{1}^{2}\frac{(b^{2}-1)}{h^{2}}%
-m_{1}^{2}}}]\text{, }
\end{equation}%
\begin{equation}
S_{2}=C_{2}\lambda _{+}[\frac{-\varepsilon _{3}\frac{b}{h}\lambda _{+}}{%
\sqrt{\lambda _{+}^{2}[\frac{b}{h})^{2}-\frac{1}{h^{2}}]-m_{3}^{2})}}-1]
\end{equation}%
Different options should be considered separately.

IN+

$C_{2}\geq 0$, $\varepsilon _{3}=-1$, so $S_{2}\geq 0$. Let, additionally, $%
bb_{2}-\frac{g_{1}}{2g_{H}}\geq 0$. Now we will show that also $S_{1}\geq 0$%
. under additional conditions which are not very restrictive. To simplify
matter further, we put $m_{3}=0$ and take $E_{3}=\lambda _{+}$ similarly to 
\cite{j}. We have

Indeed,%
\begin{equation}
\frac{\lambda _{+}^{4}}{E_{3}^{2}\frac{(b^{2}-1)}{h^{2}}-m_{3}^{2}}-\frac{%
E_{1}^{4}}{E_{1}^{2}\frac{(b^{2}-1)}{h^{2}}-m_{1}^{2}}=\frac{D}{(E_{3}^{2}%
\frac{(b^{2}-1)}{h^{2}}-m_{3}^{2})(E_{1}^{2}\frac{(b^{2}-1)}{h^{2}}%
-m_{1}^{2})}\text{,}
\end{equation}%
\begin{equation}
D=\lambda _{+}^{4}[E_{1}^{2}\frac{(b^{2}-1)}{h^{2}}-m_{1}^{2}]-E_{1}^{4}%
\lambda _{+}^{2}\frac{(b^{2}-1)}{h^{2}},
\end{equation}%
\begin{equation}
\frac{h^{2}D}{E_{1}^{4}\lambda _{+}^{2}}\equiv f(x)=x^{2}\frac{\lambda
_{+}^{2}}{E_{1}^{2}}-(b^{2}-1)\text{,}
\end{equation}%
where%
\begin{equation}
x=\sqrt{b^{2}-1-h^{2}\frac{m_{1}^{2}}{E_{1}^{2}}}\text{.}  \label{xb}
\end{equation}%
It follows from (\ref{xb}) that for $E_{1}\geq m_{1}$%
\begin{equation}
\sqrt{b^{2}-1-h^{2}}=x_{1}\leq x\leq x_{2}=\sqrt{b^{2}-1}  \label{x12}
\end{equation}%
Assuming $E_{1}=m_{1}$, we obtain from (\ref{a1b}) that%
\begin{equation}
\frac{A_{1}}{E_{1}}=\frac{b-x}{h}\text{.}
\end{equation}%
In case $m_{3}=0$, we see from (\ref{la}) that%
\begin{equation}
\frac{\lambda _{+}}{A_{1}}=h(b+\sqrt{b^{2}-1})\text{.}
\end{equation}%
Then,

\begin{equation}
\frac{\lambda _{+}}{E_{1}}=(b+\sqrt{b^{2}-1})(b-x)\text{,}
\end{equation}%
\begin{equation}
f=(b+\sqrt{b^{2}-1})^{2}(b-x)^{2}x^{2}-(b^{2}-1)\text{.}
\end{equation}%
It is seen from (\ref{xb}) that%
\begin{equation}
f(x_{2})=0\text{.}
\end{equation}%
If, additionally, we assume that%
\begin{equation}
b^{2}>\frac{4(1+h^{2})}{3}\text{,}
\end{equation}%
the derivative $\frac{df}{dx}<0$ in the range (\ref{x12}). \ In particular,
it is satisfied for the Kerr metric. Then, $f\geq 0$, so indeed $S_{1}\geq 0$%
.

If $bb_{2}-\frac{g_{1}}{2g_{H}}<0$, the quantity $S_{1}<0$, so the sign of $%
S $ can be arbitrary.

In scenarios OUT+, OUT- the quantities $C_{1}$ and $C_{2}$ can be arbitrary,
so the sign of $S$ is also arbitrary.

\end{document}